\documentclass[12pt,preprint]{aastex}
\usepackage{emulateapj5}

\begin{document} 

\title{{Halos of Spiral Galaxies. II. 
               halo metallicity-luminosity relation\altaffilmark{1}}}

\author{M. Mouhcine\altaffilmark{2,3}, 
        H.C. Ferguson\altaffilmark{4},
        R.M. Rich\altaffilmark{2}, 
	T.M. Brown\altaffilmark{4}, 
        T.E. Smith\altaffilmark{4}}
\altaffiltext{1}{Based on observations with the NASA/ESA Hubble Space 
                 Telescope, obtained at the Space Telescope Science Institute,
		 which is operated by the Association of Universities
		 for Research in Astronomy, Inc.,under NASA contract 
		 NAS 5-26555}	
\altaffiltext{2}{Department of Physics and Astronomy, UCLA, Math-Science 
                 Building, 8979, Los Angeles, CA 90095-1562}
\altaffiltext{3}{Present address: School of Physics and Astronomy, 
                 University of Nottingham, 
                 University Park,  Nottingham NG7 2RD, UK}		 
\altaffiltext{4}{Space Telescope Science Institute, 3700, San Martin Drive,
                 Baltimore, MD, 21218, USA}

\begin{abstract}
Using the Hubble Space Telescope, we have resolved individual red-giant 
branch stars in the halos of eight nearby spiral galaxies. The fields
lie at projected distances between 2 and 13 kpc along the galaxies'
minor axes. The data set allows a first look at the systematic trends
in halo stellar populations.
We have found that bright galaxies tend to have broad red-giant branch 
star color distributions with redder mean colors, suggesting that the
heavy element abundance spread increases with the parent galaxy luminosity. 
The mean metallicity of the stellar halo, estimated using the mean colors 
of red-giant branch stars, correlates with the parent galaxy luminosity.
The metallicity of the Milky Way halo falls nearly 1 dex below this 
luminosity-metallicity relation, suggesting that the halo of the Galaxy 
is more the exception than the rule for spiral galaxies; i.e., massive 
spirals with metal-poor halos are unusual.    
The luminosity-halo stellar abundance relation is consistent with the 
scaling relation expected for stellar systems embedded in dominant halos, 
suggesting that the bulk of the halo stellar population may have
formed in situ. 

\end{abstract}
\keywords{galaxies: formation -- galaxies: halos --  galaxies: stellar 
content --galaxies: individual (NGC~55, NGC~247, NGC~253, NGC~300, 
NGC~3031, NGC~4244, NGC4945, NGC~5248)}


\section{ Introduction}

It was one of the early studies of the Milky Way's stellar halo that led 
to the seminal paper on galaxy formation by Eggen, Lynden Bell, \& Sandage 
(1962).  It is not surprising that subsequently, thinking about the 
formation of stellar halo populations has been significantly driven by
investigations of the abundances and kinematics of Galactic halo stars, 
for which such a wealth of information is available. With a wealth of 
data on kinematics, chemistry, and ages, a picture is emerging in which 
the Galactic halo is a mix of stars formed very early in the Galaxy's 
formation history and stars formed later in dwarf galaxies that were
subsequently accreted. We do not know the relative balance of these 
processes, and we do not know whether the early epoch of halo formation 
was simply a time of more rapid galaxy infall or something more akin to 
a single starburst. 

While we cannot obtain the same amount of information on the halos of 
other galaxies, it is possible to resolve stars in the halos of nearby 
spiral galaxies and investigate trends. Do all spiral galaxies have halos? 
Are they metal poor or metal rich? How do the properties of the stellar 
populations correlate with other properties of the galaxies?  

The only other spiral galaxy in which the stellar halo has been resolved 
and studied in comparable level of detail to our own Galaxy is the other 
massive spiral galaxy of the Local Group, M31 (Mould \& Kristian 1986; 
Christian \& Heasley 1991; Durrell et al. 1994; Rich et al. 1996; 
Holland et al. 1996; Ferguson et al. 2002; Brown et al.\ 2003).  
The main conclusion drawn from the comparison of the Milky Way and M31 
is that the halos of the two galaxies are very different. While the 
globular cluster systems in both the Milky Way 
and M~31 halos have similar metallicity distribution functions with 
equivalent mean metallicities ([Fe/H]$\approx\,-1.6\,$dex; 
Huchra et al. 1991; Harris 1996), the metallicity distribution 
function of field halo stars in M31 is dominated by stellar populations 
considerably more metal-rich ([M/H]$\sim\,-0.5$; Mould \& Kristian 1986; 
Durrell et al.\ 2001) than observed for the Galaxy field halo stars, 
where the metallicity distribution peaks at [M/H]$\sim\,-1.5$ (e.g., 
Ryan \& Norris 1991). 
Clear indications exist that the halo of M~31 underwent more chemical 
enrichment than did the Milky Way halo (Rich et al. 1996; Durrell et al. 
2001). The surface luminosity distribution of the M31 stellar halo follows 
the de Vaucouleurs law, $R^{1/4}$ (e.g., Pritchet \& van den Berg 1994) 
out to 20\,kpc, and a $R^{-2}$ projected profile at larger distances 
(Guhathakurta et al. 2005; Irwin et al. 2005), 
in contrast with the surface density distribution in the Galaxy halo that 
follows $\rho\sim\,R^{-3.5}$ (Preston et al. 1991, Chiba \& Beers 2000; 
Vivas et al. 2001). 
Furthermore, the mean density of M31's stellar halo appears to 
be much higher than that of the Galactic halo (Reitzel et al. 1998). 
Finally, the M~31 halo shows evidence for a significant intermediate-age
population, with over half of the stars at high metallicities ([Fe/H]$>-0.5$)
and intermediate ages of 6--11~Gyr (Brown et al.\ 2003), in striking
contrast to the ancient metal-poor halo of the Milky Way.
It is not clear how these differences may be related to different halo 
formation histories, and/or to different initial conditions in the 
protogalaxies. We do not know whether the properties of the Milky Way halo 
are the rule or the exception for spiral galaxy halos.

A key quantity that helps disentangle the formation history of the stellar 
halo is the galactic halo stellar metallicity.
Spectroscopy of giant stars in halos is difficult and time consuming, even
for stars within the halos of Local Group galaxies (see Reitzel \& Guhathakurta 
2002 for the first spectroscopic observations of giant stars in M31 halo). 
Fortunately the colors of red giant stars on the color-magnitude diagram 
are predominantly sensitive to metallicity rather than age, so a 
first-approximation estimate of the stellar metallicity can be derived 
through stellar photometry.

Our knowledge of the metallicities of stellar populations in the outskirts 
of galaxies is improving rapidly thanks to wide field cameras on the ground 
and sensitive observations with HST. For the handful of galaxies studied to 
date, the halo stellar populations are generally higher in metallicity than 
the Milky-Way halo. 
Early imaging in the M31 halo by Mould \& Kristian (1986) revealed a metal
rich red giant branch. The high resolution imaging capability of HST made more 
halo populations accessible:  Soria et al. (1996) and Harris \& Harris (2000) 
find a wide metal rich giant branch in NGC 5128, and Elson (1997) measured a 
wide and metal rich giant branch in the halo of NGC 3115. Even in the dwarf 
galaxy NGC~147 a wide and relatively metal rich red giant branch is seen 
(Han et al. 1997). Yet up to now, no systematic survey has been undertaken 
using HST to characterize the properties of Galactic halo field stars as a 
function of parent galaxy properties.

Using photometry of resolved giant stars in the halos of a sample of edge-on 
spiral galaxies, we obtain the red giant branch locus and thence an inference 
on metallicity. We report the discovery of a scaling relation between the mean 
stellar metallicity of halo field stars and the luminosity of the host galaxy. 

\section{ The Data}
\label{data}
 
\begin{figure}
\includegraphics[height=3.in]{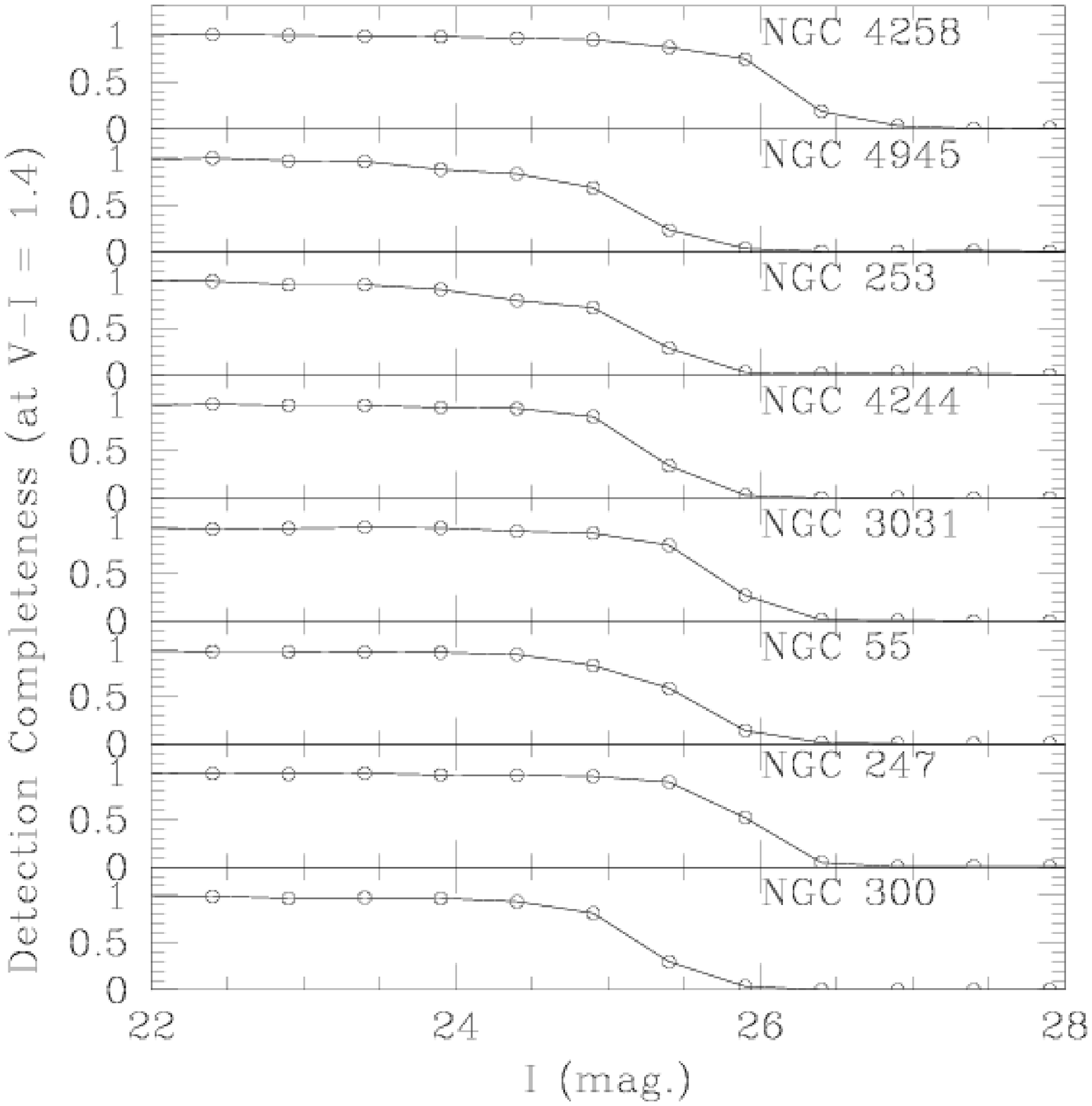}
\includegraphics[height=3.in]{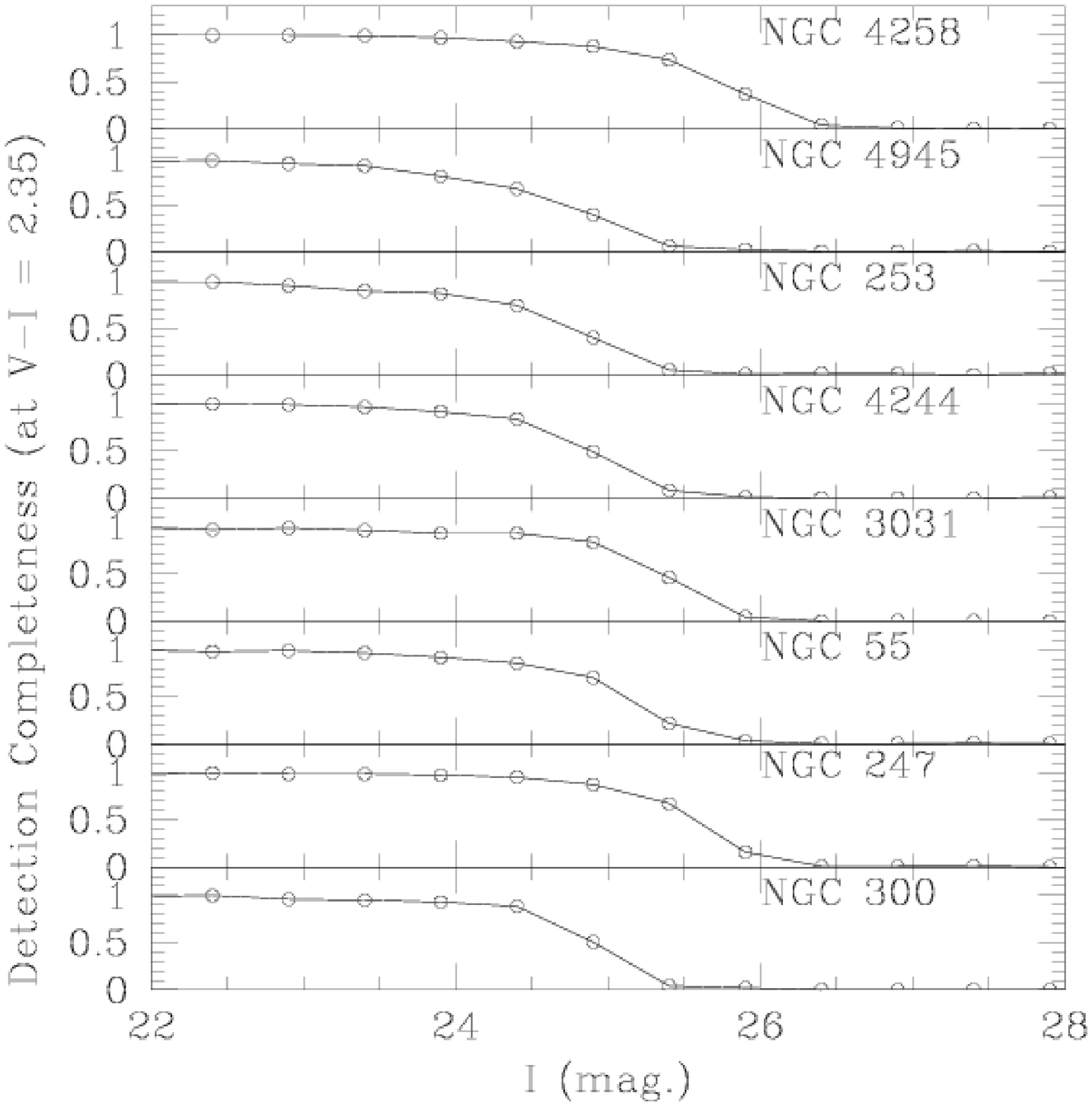}
\caption{Detection completeness as a function of magnitude 
for each of the WFPC2 fields, as determined from artificial-star 
experiments, for (V-I)=1.4 ({it Left}), and (V-I)=2.35 ({it Right})
respectively. }
\label{compl}
\end{figure}

The galaxy selection criteria, observations, and the reduction techniques 
applied to our sample of highly inclined spiral galaxies are described 
by Mouhcine et al. (2005a). The locations of the observed halo fields,
superimposed on the Digitized Sky Survey images of these galaxies, are 
shown in paper I and paper III of this series. The integrated magnitude, 
the foreground extinction-corrected $(J-K)_{\circ}$, and the distance 
modulus of the galaxies in our sample are listed in Table~\ref{gal_prop} 
(see paper III of this series for more details on the adopted distance 
moduli for the sample galaxies). Extinction- and inclination-corrected 
V-band magnitudes come from de~Vaucouleurs et al. (1991; RC3). 
The infrared photometry is taken from the Two Micron All-Sky catalogue 
(Jarret et al. 2000). No significant internal extinction is expected 
to affect the halo stellar populations, as halo regions are most likely 
dust free. The reddening toward each halo field was estimated using the 
all-sky map of Schlegel et al. (1998). 
The images are sufficiently deep to reveal RGB stars with roughly a 
solar metallicity in almost all the observed fields, but in some cases 
it requires careful modeling of incompleteness to characterize the 
high-metallicity end. Fig.\,\ref{compl} shows the completeness levels 
as a function of I-band magnitudes for the observed halo fields, 
determined from artificial-star experiments (see paper III of this 
series for more details on artificial star experiments and completeness 
levels of the data). The completeness levels are shown for the particular 
colors (V-I)=1.4 (left panel) and (V-I)=2.35 (right panel).
Due to observing time constraints, the depths of the images vary from 
galaxy to galaxy, with 50\% completeness limits ranging from absolute
magnitudes $-0.9$ to $-3.5$. 

Our use of the term ``halo'' throughout this paper is necessarily imprecise.
While our fields are projected along the minor axes well outside the visible
outskirts of the disks and bulges on the Palomar Sky survey, there may 
nevertheless be some bulge, disk and/or thick disk stars in our fields. 
It might be possible to quantify the possible contribution of these 
components by fitting parametric models to deep near-infrared surface
photometry, but we do not have such data for the full sample, and the
definition would be largely rely on the extrapolation of these parametric
fits well beyond where they are constrained by photometry. We note that
even in the well-studied case of M31, where the outer galaxy can be studied
by counting stars, it is difficult to define where the bulge or disk ends 
and the halo begins. Thus when we refer to ``halo'' we simply mean ``stars
along the minor axis, well outside the visible bulge and disk.''

\section{ Halo stellar metallicity-luminosity relation}
\label{halo_lz}

\begin{figure}
\includegraphics[height=3.5in]{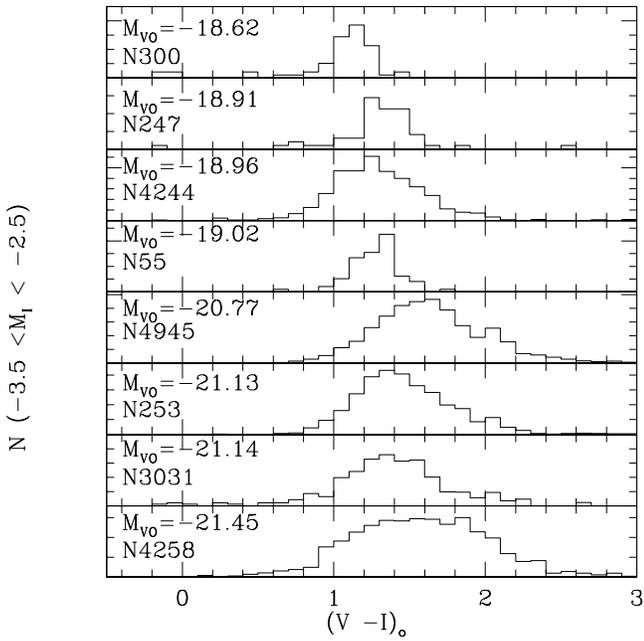}
\caption{Incompleteness- and extinction-corrected color 
distribution of stars with absolute magnitudes 
$-3.5\le\,M_{I}\le\,-2.5$.The galaxies have been arranged 
in order of absolute face-on V-band magnitude, corrected 
for both Galactic extinction and galaxy inclination.}
\label{galcdf}
\end{figure}

A detailed discussion of the color-magnitude diagrams of our 
sample galaxies is presented in Mouhcine et al. (2005b), to which the reader 
is referred. The halo stellar populations of spiral galaxies are found to be
dominated predominantly by red giant stars, indicating stellar populations
older than 1~Gyr (e.g., Sweigart, Greggio, \& Renzini 1990). 
It is well known that the colors of giant stars in an old simple stellar 
population are mainly affected by the abundances of heavy elements and only 
to a much smaller extent by the age of the stellar population.  For example, 
at $M_I = -3$ on the red giant branch, a 0.2~mag shift in $V-I$ to the blue 
can be achieved by a relatively large decrease in age from 13 to 6 Gyr or a 
relatively small decrease in [Fe/H] from $-1$ to $-1.2$.  Thus, at a given 
luminosity, redder stars are generally more metal-rich than bluer ones, and 
the width of the red giant branch at a given luminosity is an indicator of 
the spread in the stellar metallicity. Fig.\,\ref{galcdf} shows the 
incompleteness- and foreground-extinction corrected color distributions of 
red giant branch stars within the magnitude range $-3.5 \le M_I\le -2.5$ 
for the galaxy sample ranked by luminosity. 

Our photometry for NGC 4258 does not reach as deep relative to the
RGB tip as it does in other galaxies in the sample. It is clear 
from the photometry at the bright end of the RGB that the metallicity
distribution is broad and extends to high metallicity. However, the 
50\% completeness level is sufficiently bright (see Fig. 1) that it 
makes our interpretation of the high-metallicity end of the distribution
highly uncertain. The effect of incompleteness was included in our
construction of the color distribution to the best of our ability,
but the corrections become large and uncertain at the highest
metallicities. If we have underestimated our incompleteness, the
color distribution for NGC4258 would move to the red and the width
would broaden. Further deep observations in the optical and near-IR
of this galaxy are clearly warranted. However, we should mention 
that the conclusion drawn in this paper do not rely on NGC~4258.

Considering the color distribution of giant stars alone, two striking 
properties emerge. First, the mean color tends to be redder for 
brighter galaxies, and second, the width of the color distribution 
tends to be wider for brighter galaxies. We see no evidence for 
bi- or multi-modality in the color distribution of giant stars; 
however this does not necessarily mean that the metallicity 
distributions are unimodal, because the relation between the color 
and metallicity of giant stars is nonlinear.

Once the stellar photometry is extinction- and distance-corrected, 
we can estimate the mean metal abundances via a comparison with 
the fiducial globular cluster giant branches. Lee et al. (1993) 
have provided a quadratic abundance calibration based on the mean 
$(V-I)_{\circ}$ color of the red giant branch at a luminosity of 
$M_I = -3.5$ for the full abundance range of the calibration 
clusters (${\rm -2.2\le [Fe/H]\le -0.7}$). Adopting $M_I = -3.5$~mag 
as a reference for the abundance determination increases the sensitivity 
to abundance, minimizes the influence of asymptotic giant branch stars, 
and also reduces the effect of photometric errors in most cases. 
Metal-rich stars, i.e., ${\rm [Fe/H]\ga\,-0.5}$, lie below $M_I = -3.5$.
No calibration of the relationship between stellar metallicity and 
(V-I) color at a fainter absolute I-band magnitude is available for 
such metallicities. If metal-rich stars are present in the halos 
of spiral galaxies, they will be not taken into account to calculate 
the mean stellar halo metallicity using Lee et al. (1993) calibration,
and then possibly biasing the estimated metallicities toward low 
metallicities. However, the color-magnitude diagrams of the observed 
halo field, presented in paper III of this series, show that while 
a metal-rich stellar population is present, it is not dominant (see 
paper III of this series for a detailed analysis of the metallicity 
distribution functions for the galaxy sample). Thus we do not
expect any significant bias toward low metallicities.
To measure the ${\rm (V-I)_{\circ,-3.5}}$, we calculate the histogram 
of the colors of stars within $M_I = -3.5\pm 0.1$. To determine the 
uncertainties, we carry out a bootstrap resampling procedure. For each 
simulated sample, the histogram of stellar colors is fitted by a 
Gaussian. We then fit a Gaussian to the final distribution of 
${\rm (V-I)_{\circ,-3.5}}$ for the simulated samples, and use its mean 
and dispersion as the best estimate of ${\rm (V-I)_{\circ,-3.5}}$ and 
its uncertainties. We have repeated the same procedure using the median 
value for the simulated samples instead of the peak of the Gaussian, 
or using a mean value of simulated samples means, and the differences 
never exceed 0.02 dex.
The estimated mean ${\rm (V-I)_{\circ,-3.5}}$ color for the 
observed halo field of NGC~4945 is slightly redder than the color
of the most metal-rich globular cluster used to calibrated the
${\rm (V-I)_{\circ,-3.5}}$ vs. [Fe/H] relation, but still consistent 
within the errors. To estimate the mean stellar halo metallicity for 
NGC~4945, we have simply extend Lee et al. (1993) calibration to the 
observed mean ${\rm (V-I)_{\circ,-3.5}}$ color. 

Note that because of the low completeness level at $M_I = -3.5$~mag for 
NGC~4258 halo red stars, i.e., ${\rm (V-I)\ga\,2}$, we are possibly missing 
a fraction of metal-rich/red stars. The mean $(V-I)_{\circ}$ color of halo 
red giant branch stars at a luminosity of $M_I = -3.5$ for this galaxy might 
be biased toward a bluer color. The estimated mean metallicity of NGC~4258 
halo stars might be then underestimated. The errors are the formal errors 
of the mean; the overall uncertainty in the mean metallicities is 
conservatively estimated to be $\pm 0.3$ dex because of systematic errors 
in the photometry and/or the calibration. In paper III, we compute 
metallicity distributions by applying corrections star by star and obtain 
consistent results.

\begin{figure}
\includegraphics[height=3.5in]{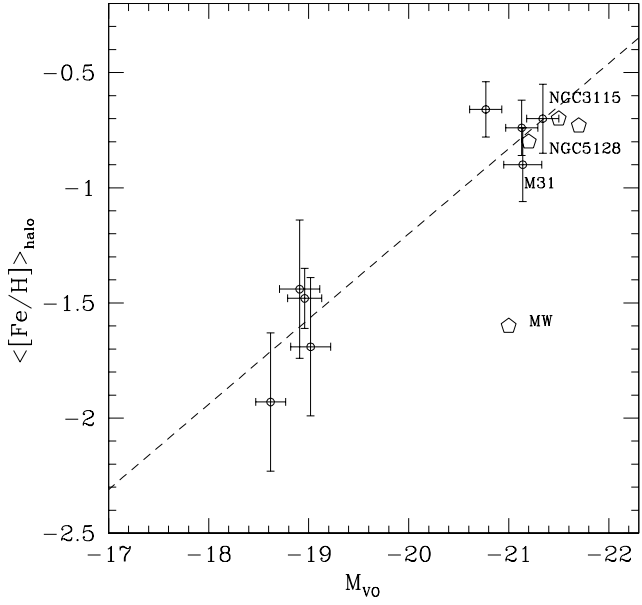}
\caption{Mean metallicity of halo stellar populations plotted 
against parent galaxy absolute luminosity. Pentagons display 
data collected from the literature for galaxies named in the 
figure. MW indicates Milky Way galaxy.
The dashed line is a rough fit to the [Fe/H]-M$_{V}$ 
relation, $L\propto Z^{2.7}$, similar to the [Fe/H]-M$_{V}$ 
relation of Dekel \& Silk (1986) for dwarf stellar objects 
formed in potential wells.}
\label{zl_halo}
\end{figure}

Fig.\,\ref{zl_halo} shows the variation of the mean abundance of the halo 
field stars as a function of the absolute V-band magnitude of the parent 
galaxy for our galaxy sample. In order to examine the luminosity-stellar 
halo metallicity relation over a large range of luminosity and metallicity, 
we have augmented our database of halo field star mean abundances with 
published measurements for galaxies observed to sufficient depths to measure 
securely the halo star abundances. We have overplotted the values of the 
stellar halo abundances of M~31, the Milky Way, the giant E/S0 NGC~5128, 
and the S0 galaxy NGC~3115. 
All investigations of the M~31 stellar halo, following the pioneering work 
of Mould \& Kristian (1986), find the dominant field population has a mean 
metallicity of around ${\rm [Fe/H]\sim -0.8}$ dex, over a wide range of 
projected distances from $\sim 5$~kpc to $\sim 30$~kpc (Rich et al. 1996;    
Holland et al. 1996; Durrell et al. 2001, 2004; Bellazzini et al. 2003).
A large observational effort was dedicated to investigate the abundance 
distribution of the Galaxy halo. Both kinematically selected (Laird et al. 
1988; Ryan \& Norris 1991) and non-kinematically selected (Beers et al. 2000) 
halo star samples show that the metallicity distribution function peaks 
at ${\rm [Fe/H]\sim -1.5}$ dex. The galaxy luminosities are taken from the 
Local Group catalog of van den Bergh (2000). Harris \& Harris (2000, 2002) 
have shown that the stellar halo of NGC~5128 is dominated by a metal-rich 
component extending to a solar metallicity with a mean metallicity of 
${\rm [Fe/H]\approx -0.7}$ dex. Kundu \& Whitmore (1998) have reanalyzed 
deep imaging data for NGC~3115 from Elson (1997), and published the color 
distribution of bright halo giant stars. Using a distance modulus of 
$(m-M)_{\circ}=30.2$, fitting a Gaussian to the color distribution in the 
magnitude range $-3.7<M_I<-3.5$ (as given in Kundu \& Whitmore 1998), and 
converting the mean color to a mean metallicity using Lee et al. (1993)
calibration, we found that the mean halo metallicity is $[Fe/H]\approx -0.7$.
The metallicity of the stellar populations in the outer parts of M~33 
have been measured. However, we have not included M~33 to define the stellar 
halo metallicity-luminosity relation, as it is not clear to which galactic 
component, i.e., galactic disk or stellar halo, stars in the outer parts of 
M~33 belong to. Mould \& Kristian (1986) have found that the stellar 
halo of M~33 is dominated by metal-poor stars, i.e., ${\rm [Fe/H]\sim -2}$ 
dex, with a small spread. Tiede et al. (2004) have imaged a field in the
outer part of M~33 including the Mould \& Kristian (1986) field but find 
${\rm [Fe/H]\sim -1}$ and an age gradient. They argure that they are in 
fact imaging the outer disk of M33 (see also Brooks et al. 2004 for similar 
results). The metallicity gradient within their observed field is consistent 
with the spatial variation of stellar metallicity seen in the inner disk 
regions of M~33, suggesting that the majority of stars in their field 
belong to the disk, not the halo (Tiede et al. 2004). Recent observations 
show that stars  in the outer parts of M33 are distributed exponentially 
as within discs (R. Ibata, private communication). The outer part of M33 
seems to be dominated by disk stars rather than halo stars. 
Most of the galaxies in our sample with types and luminosity similar
to M33 are highly inclined (apart from NGC~300). Thus our fields
are much more likely to be dominated by halo stars than any field
yet observed in M33.
%
Fig. 2 shows that a good correlation is present in the data between the 
parent galaxy luminosity and the mean abundances of field halo stellar 
populations. 

For comparison, the dashed line indicates the scaling relation, 
${\rm L \propto Z^{2.7}}$, expected for objects originating as gaseous
protogalaxies embedded in dominant dark matter halos framed within the 
context of a single star formation event, and whose the chemical 
enrichment was dictated by enrichment from massive stars and gas loss 
via supernovae-driven winds under the assumption of instantaneous mixing 
(Dekel \& Silk 1986). The slope of the predicted luminosity-metallicity 
relation is based on the key assumption that the galactic objects are 
embedded in dominant dark matter halos. Note the arbitrary metallicity 
zero point of this relation. 

Surprisingly enough, both early type galaxies, i.e., NGC~5128 and NGC~3115, 
fall on the relation defined by spiral galaxies, 
suggesting that the diffuse halo stellar populations in elliptical galaxies 
may share some similarities with spiral halo stellar populations. 
One possibility is that in most cases we are observing the outer extensions 
of the bulge/spheroid. One may also consider the possibility that giant 
early-type galaxies might have experienced a chemical enrichment and star 
formation history in their outer regions similar to that seen in the halos 
of typical bright spiral galaxies.
Larsen et al. (2001) have recently analyzed the properties of globular cluster 
systems around a large sample of early-type galaxies, and their correlations as 
functions of the parent galaxy properties. They concluded that the data appear 
to favor a scenario where globular cluster systems as a whole form in situ after 
the elliptical progenitor has been assembled into a single potential well. 
In both scenarios, where halo globular clusters and field stars form roughly 
contemporaneously, or where halo field stars are the results of tidal stripping 
of the globular clusters, one may expect that field halo stellar populations in 
ellipticals should obey the luminosity-halo stellar abundance relation of spiral 
galaxies. Harris \& Harris (2000, 2002) have shown that it is unlikely that
halo of NGC~5128 was built by accretion of pre-existing, gas-free small satellite 
galaxies, and favor an in situ formation model where the star formation proceeds 
in two stages.
To investigate in more detail the similarities between spiral galaxy halos and 
early-type galaxy halos, data of the necessary depth for other normal elliptical 
galaxies, such as NGC~3379, are needed. 
   
The most striking result from the figure is that the mean metallicity of the 
Milky Way stellar halo does not fit onto the luminosity-halo stellar metallicity 
relation, lying low by an order of magnitude. This suggests 
that the stellar halo of the Galaxy {\it may not be typical for a normal spiral 
galaxy of its luminosity}, i.e., galaxies with luminosities similar the Galaxy 
luminosity clustered at $[Fe/H]\sim -0.7$ compared to $[Fe/H]_{MW}\sim -1.6$. 
Clearly a larger sample is needed to determine what parameters, other
than total luminosity, correlate with halo metallicity.

Recently, Bekki et al. (2003) have investigated numerically the physical 
processes that may produce the observed metallicity distribution function 
of NGC~5128, considering a model in which the galaxy has been formed by the 
merging of two spiral galaxies. One of the key inputs of their simulations 
is the metallicity distribution of the stellar halo of a merger progenitor 
spiral. They consider that the field halo stellar population of a typical 
spiral galaxy is similar to what is observed for the Galaxy, in contradiction
with our finding. To populate the halo of the merger product with metal-rich 
stars, as is observed for NGC~5128, their models require that the stellar halo 
be composed of stars that were located in the outer parts of the two merging 
disks as they were tidally stripped during the merging event. If halos are in 
general more metal rich, it might make this model less tenable.

The figure shows that all of the sample galaxies fall along the 
$L \propto Z^{2.7}$ scaling relation. Unfortunately, the sample galaxies
clustered into two clumps.  Intermediate luminosity galaxies, i.e,
$M_{V,\circ}\sim -20$, are missing in our sample, so we cannot draw a firm
conclusion about their location in the stellar halo luminosity-metallicity
relation.  Furthermore, galaxies brighter than the brightest in our sample
are suspected to behave differently. Indeed, Kauffmann et al. (2003)
have shown that galaxy properties change dramatically at stellar
masses of $\sim 3\times 10^{10} M_{\sun}$; galaxies with stellar
masses above this limit have a larger fraction of old stellar
populations, and concentrations typical of bulges, while galaxies
with lower stellar masses are dominated by relatively young stellar
populations and have low concentrations typical of disk galaxies.
This behavior contrasts with the evolution of dark matter halo
properties, which should vary smoothly with mass. It will be important 
to measure the metallicities of halos of galaxies with intermediate 
luminosities to further explore the connection between halos and their 
parent galaxies. Nevertheless, the apparent fact that stellar halos 
follow a luminosity-metallicity relation similar to what is predicted 
for gas clouds embedded in massive halos is consistent with a halo
formation scenario where field stars formed in the virialized potential 
well of the parent galaxy.  
As mentioned in the introduction, it is likely that our fields
contain a mix of stars traditionally associated with halo, bulge
and thick disk. Some of the galaxies in our sample have prominent 
bulges (e.g. NGC~4258 and NGC~4945), and it is possible that our 
stellar samples in these galaxies are dominated by the extension 
of the bulge component. The contribution of bulge stars could 
bias the mean stellar halo metallicity toward higher metallicities. 
If this is true, the variation of galaxy properties, e.g., 
bulge-to-disc ratio, at a given galaxy luminosity might produce 
a variation of mean stellar halo metallicities for galaxies with 
similar luminosities.
The observed kinematics of M~31 halo stars support this possibility. 
Indeed, they are rapidly rotating similar to the M~31 bulge 
(Hurley-Keller et al. 2004; see Perrett et al. 2002 for a similar 
result for the kinematics of globular cluster system around M~31). 
Thus our observed halo metallicity-luminosity relation could be 
indicating the trend for more luminous galaxies to have more 
prominent bulges. It is unclear how this relates to the merger 
vs. in-situ star-formation scenarios, since both bulges and 
halos can grow by either process. A key test will be to obtain
metallicities of extra-planar stars in massive late-type 
(e.g. Sc) galaxies without prominent bulges.

Alternatively, halos may form primarily by the disruption of 
relatively gas-free dwarf galaxies orbiting around the parent galaxy. 
If the stellar halos were formed hierarchically from the disruption 
of dwarf galaxies, one might expect that the stellar halo of 
brighter galaxies may be dominated by metal-rich stars as the mass 
function of galactic satellites around bright galaxies might extend 
to larger masses.
Unfortunately, the slope of the metallicity-luminosity relation in
such scenarios is not easily predictable, and may depend on a large
number of unknown ingredients, e.g., dwarf galaxy mass function,
feedback at low mass scale, and the detailed merging history. 
It will require more detailed modeling to determine whether a pure 
accretion model such as this can be ruled out. 

\begin{figure}
\includegraphics[height=3.5in]{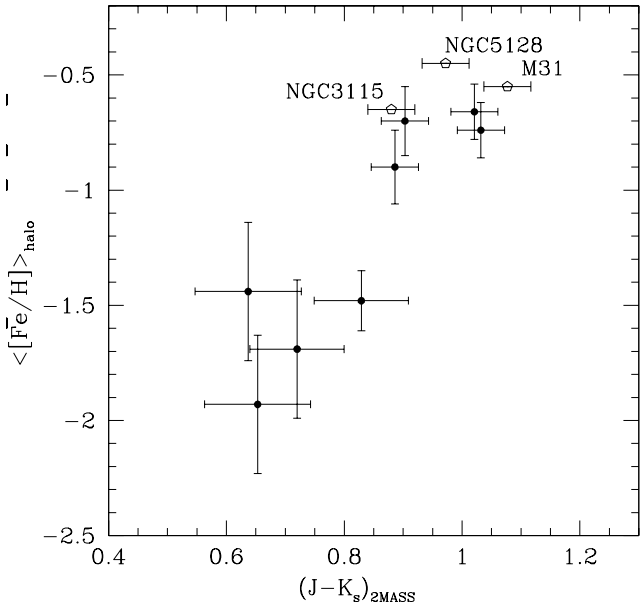}
\caption{Mean metallicity of halo stellar populations plotted 
against the parent galaxy foreground extinction-corrected (J-K).}
\label{zcol_halo}
\end{figure}

\begin{figure}
\includegraphics[height=3.5in]{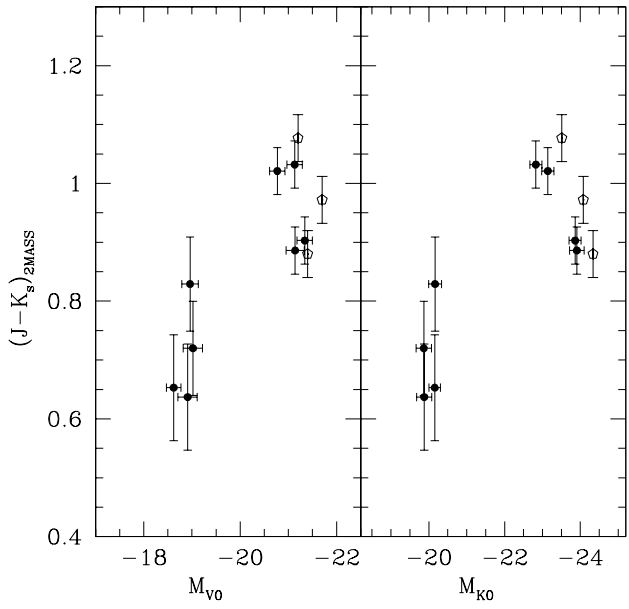}
\caption{Color-magnitude relation for the sample galaxies. The colors are
calculated using integrated magnitudes.}
\label{cmr_sample}
\end{figure}

Fig.\,\ref{zcol_halo} shows the evolution of the halo field star
mean metallicity as a function of the integrated foreground
extinction-corrected near-infrared color $(J-K)_{\circ}$. The stellar
halo mean metallicity correlates with the parent galaxy color. Taken
at face value, this may indicate that the star formation and chemical
evolution histories of both the stellar halo and the body of the
parent galaxy may be connected. This comes as a surprise, since,
as it is well established, disk galaxies are, generally speaking,
still actively forming stars, while stellar halos are not.
What may cause a connection between the
halo metallicity and the parent galaxy color? For the galaxy sample,
augmented by data from the literature, $(J-K)_{\circ}$ correlates
with the galaxy luminosity as shown in Fig.\,\ref{cmr_sample}, with
rather a large scatter.  Combined with Fig.\,\ref{zl_halo}, the
apparent correlation in Fig.\,\ref{zcol_halo} is driven by the
color-magnitude correlation of the sample galaxies. The sample
galaxies also follow an optical-infrared color-magnitude relation.

It is well-established that early-type galaxies obey a tight
color-magnitude relation (de Vaucouleurs 1961, Sandage \& Visvanathan
1978, Bower et al. 1992).  This correlation is generally understood
as a change in metallicity as the galaxy mass increases, i.e., a
metallicity-luminosity sequence (Vazdekis et al. 2001).  A similar
relation is observed for spiral galaxies (Tully et al. 1982; Wyse
1982).  This is generally interpreted as a combined effect of increasing
metallicity (Bothun et al. 1984), and/or of a decreasing fraction
of blue stars (Peletier \& de Grijs 1998) with increasing galaxy
mass, with no clear sensitivity to the Hubble type (Gavazzi 1993).
The issue is still a matter of debate, because of the effects of
both complex star formation and chemical evolution histories for
spiral galaxies, as well as the internal reddening (see Arimoto \&
Jablonka 1991 for a discussion).

The $(J-K)_{\circ}$ vs. $M_{H}$ relation for our sample is in good
agreement with the disk dominated spiral galaxy relation presented
by Bothun et al. (1984). Bright galaxies are expected to be more
affected by internal reddening than fainter ones (Jansen et al.
2001; Mouhcine et al. 2005c). For the sample galaxies, the change
of infrared colors with absolute magnitude is small, and differential
reddening of $\delta E(B-V) \sim 0.3$, can account for a color
offset of $\delta (J-K)\sim 0.2$, similar to what is observed. The
short baseline of infrared colors does not allow a good continuum
definition. The color-magnitude relation shows up again in the
optical-infrared regime: the (V-K) color is bluer for fainter
galaxies, in a sense that cannot be account for by a simple
differential reddening between faint and bright. So, the galaxy
mass/luminosity regulates, in broad terms, the stellar content of
spiral galaxies. Thus the correlation seen in Fig.\,\ref{zcol_halo}
between the halo field star mean metallicity and the galaxy color
is not indicative of any fundamental evolutionary link between the
halo and the galaxy disk, other than the fact that the evolution
of both disk and halo stellar populations is dominated by the galaxy
luminosity.  

\section{Summary \& Conclusions}
\label{concl}

Using a data set from HST imaging of spiral galaxies, we have shown
that the colors of halo red-giant branch stars correlate with parent
galaxy luminosity. As galaxy luminosity increases, the color
distribution broadens and the peak moves to redder colors.  We have
interpreted this trend in color as a trend in the mean metallicity of
the halo populations, using color-metallicity relations calibrated on
the Galactic globular clusters, and find
that the mean abundance of halo field stars correlates with the
galaxy luminosity.  Galaxies with accurate determinations of the
stellar halo mean abundances follow the new luminosity-metallicity
relation, with the exception of the Milky Way halo, 
which is almost an order of a
magnitude more metal-poor than galaxies with similar luminosity.
Thus the Milky Way may not have a typical halo.  Its stellar
population differs substantially from those found in the
halos of other bright spiral galaxies.  Our finding reinforces the
view that for large galaxies of all types, moderately metal-rich
stellar halos should be the norm, not the exception.  The slope
of the luminosity-metallicity relation suggests that the bulk
of the diffuse stellar halo forms in the gravitational potential
of the final galaxy, and was not formed in dwarf-like galaxies that
were tidally stripped subsequently.

\acknowledgments

M.M would like to thank Mike Merrifield for critical reading of an 
earlier version of the draft. We acknowledge grants under HST-GO-9086 
awarded by the Space Telescope Science Institute, which is operated 
by the Association of the Universities for Research in Astronomy, Inc., 
for NASA under contract NAS 5-26555

\clearpage

\begin{deluxetable}{lccccc}
\tablewidth{3.5in}
\tablecaption{Basic properties of the sample galaxies. Columns: (1) Galaxy 
name; (2) Galactic foreground extinction-corrected V-band absolute magnitude; 
(3) $(J-K)_{\circ}$ color corrected for Galactic foreground extinction; 
(4) distance modulus; and (5) mean metallicity of the stellar halo.}
\tablehead{
\colhead{Galaxy}                & 
\colhead{$M_{VO}$}              & 
\colhead{$(J-K)_{\circ}$}       &
\colhead{$(m-M)_{\circ}$}        & 
\colhead{$<[Fe/H]>_{\rm Halo}$}  }
\tablecolumns{4}
\startdata
NGC 55    &  -19.02 &  0.72 & 26.11 & -1.69 \\ 
NGC 247   &  -18.91 &  0.64 & 27.30 & -1.44 \\ 
NGC 253   &  -21.13 &  1.03 & 27.59 & -0.74 \\ 
NGC 300   &  -18.62 &  0.65 & 26.53 & -1.93 \\ 
NGC 3031  &  -21.14 &  0.89 & 27.80 & -0.90 \\ 
NGC 4244  &  -18.96 &  0.83 & 27.88 & -1.48 \\ 
NGC 4258  &  -21.30 &  0.90 & 29.32 & -0.70 \\ 
NGC 4945  &  -20.77 &  1.02 & 27.56 & -0.66 \\ 

\enddata
 
\label{gal_prop}
\end{deluxetable}

\clearpage

\end{document}